\documentclass[pdftex,11pt,twoside]{article}
\usepackage{./asp2014}
\usepackage{titlesec}
\aspSuppressVolSlug
\resetcounters

\DeclareSymbolFont{matha}{OML}{txmi}{m}{it}% txfonts
\DeclareMathSymbol{\varv}{\mathord}{matha}{118}

\titlespacing*{\section}{0pt}{1.5ex plus .1ex minus .1ex}{2.3ex plus .2ex}

\begin{document}

\title{Time evolution of Hanle and Zeeman polarization in MHD models}
\author{E.S. Carlin,$^1$
\affil{$^1$Istituto Ricerche Solari Locarno, 6600, Locarno,  Switzerland\\
\email{edgar@irsol.ch}}
}

% This section is for ADS Processing.  There must be one line per author.
\paperauthor{E.S. Carlin}{edgar@irsol.ch}{0000-0002-0012-6581}{Istituto Ricerche Solari Locarno}{}{Locarno}{Ticino}{6600}{ Switzerland}

\begin{abstract}
Exposing the polarization signatures of the solar chromosphere
requires studying its temporal variations, which is rarely
done when modelling and interpreting scattering
and Hanle signals. The present contribution sketches the scientific problem of
solar polarization diagnosis
from the point of view of the temporal dimension, remarking some key
aspects for solving it. Our time-dependent calculations expose the
need of considering dynamics explicitly when modelling and observing
scattering polarization in order to achieve effective
diagnosis techniques as well as a deeper knowledge of the second solar spectrum.
\end{abstract}

\section{Introduction}
\begin{figure}[h!]
\centering
\includegraphics[scale=0.5]{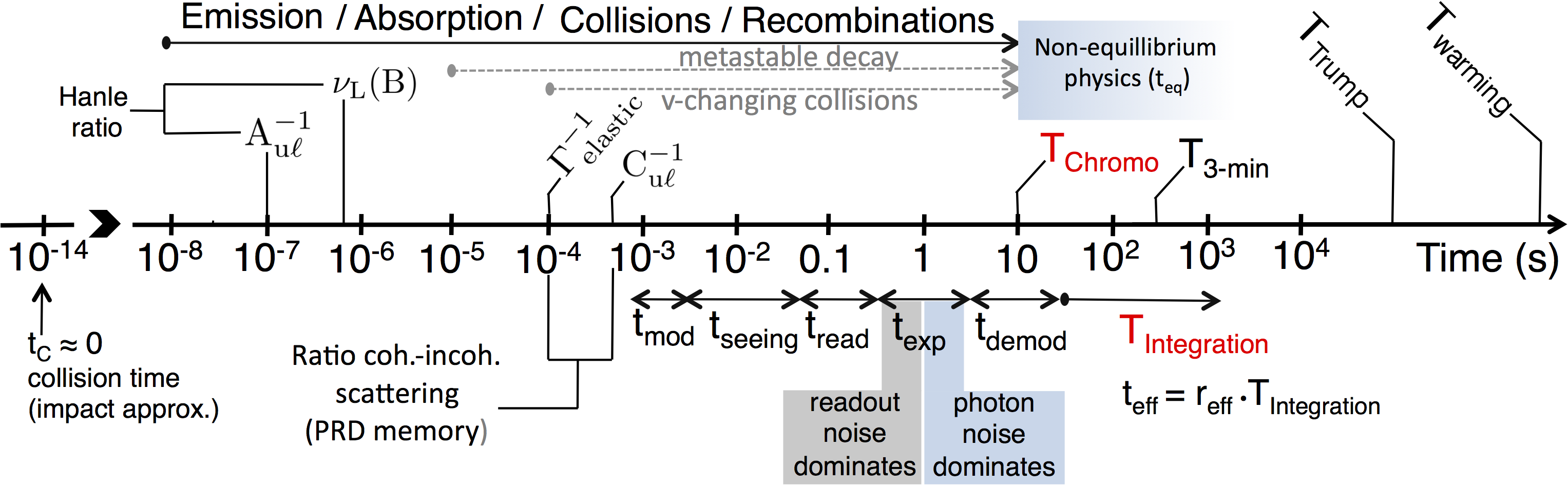} 
\caption{\footnotesize Temporal scales explaining chromospheric
  Non-LTE polarization \protect\footnotemark.}
\label{fig:timeline}
\end{figure}

Understanding the 
 scattering polarization generated by a stellar atmosphere in Non-LTE
 depends on how accurately is the temporal evolution considered. An
 example of the central role of time in such a context is the mere fact that the radiative transfer
 equation (RTE) and the rate equations for polarized light result from applying
 the Schrödinger equation to the time-evolution operator
 \citep{Landi-DeglInnocenti:1983, Bommier:1997aa}. In the solar case,
 time evolution plays a role over a huge range of scales (see Fig. \ref{fig:timeline}). The smallest time scales, between $10^{-14}$
 and $1$ s, are associated to quantum processes of interaction
 between matter and matter (collisions), matter and
 radiation (absorption and emission), and matter and magnetic field (Hanle
 and Zeeman effects), either with atoms that preserve partial
 temporal memory of the incident light during scattering (partial
 redistribution, PRD) or that do not (CRD). In medium
 scales ($10^{-2}<\Delta \rm t(s) < 10^{2}$), the response of the detectors
 is critical, particularly in terms of temporal integration, and of
 management of noise sources (seeing, readout, shot, thermal)
 peaking at different characteristic times during detection. 
 Finally, in scales $> 10$ s, chromospheric motions  
have a key impact in the generation and transfer of polarized light. 

 The temporal perspective of the
aforementioned processes suggests that their very
different time-scales can decouple them, allowing a division in
smaller subproblems. Thus, an explicit care of the temporal dimension
is substituted by %standard
generally good approximations that are universally
adopted (e.g., statistical equilibrium). However, when processes
overlapping in time are modelled separately, or when a quantity is not
 well resolved, inconsistencies and paradoxes may occur.
A trivial example of this is the integration of chromospheric dynamic
signals whose characteristic time scales are significantly shorter than the integration
time. If in addition chromospheric polarization
is modelled in absence of temporal evolution, the theoretical
predictions change radically. Recently,
\cite{Carlin:2016aa} investigated these issues by carrying out the first detailed simulation of 
temporal evolution of Hanle and Zeeman polarization, done for the Ca {\sc i}
$4227$ {\AA} line with help of
chromospheric MHD models
\citep{Carlsson:2016aa}. For more details on the calculations, see Carlin et al. (2017,
submitted). Next sections provide some contextual explanations and illustrative results.
\footnotetext{ In this SPW8 we have seen how powerful are neural
  networks. To this regard, Fig. 1 shows the result of a curious
  experiment. After training
 a neural network with our calculations and with contextual
 data, the output is found to depend on the initial political
 conditions. Training with BBC news of 2015, the network predicts
 human extinction by global warming in $\rm \Delta t>T_{warming}$. But,
 including news about the last U.S. presidential
  elections, most outputs are \$ and extinction occurs in $\rm T_{\rm
    Trump}<<T_{warming}$ due to stimulated nuclear fission, 
 patriotism of mass destruction or lack of convergence.}
\section{Theoretical considerations about dynamics}
The solution of the Non-LTE RT problem with atomic polarization
implies to calculate the optical properties of the plasma. This is
done by solving the rate equations, which describe the evolution of the atomic
density matrix $\rho^K_{Q,\mathrm{i}}$ in each atomic level $i$ 
%of angular momentum $\rm{J_i}$ 
considered \citep{LL04}. Symbolically, the comoving-frame rate equations for a plasma element 0 moving with velocity $\vec{\upsilon}_{\mathrm{macro}}$ are of the kind:
\begin{equation}\label{eq:ssev}
\centering 
\frac{\partial\rho^K_{Q,i}}{\partial t}+\vec{\upsilon}_{\rm macro}\cdot
\vec{\nabla}\left[ \rho^K_{Q,i}\right]=\,
f^{\nu_L}_{B}\left(\vec{B},\rho^K_{Q,i^{\prime}} \right) +
f^{A_{u\ell}}_{\rm
  rad}\left(J^K_Q,\rho^K_{Q,i^{\prime}}\right)+f^{C_{ii^{\prime}}}_{\rm col}\left(n,T,\rho^K_{Q,i^{\prime}}\right)
\end{equation}
Here, the total temporal variation of the atomic collectivity in 0 has magnetic ($f_{\rm B}$), radiative
($f_{\rm rad}$)
and collisional ($f_{\rm col}$) contributions whose maximum values
have representative orders of magnitude
given by the Larmor frequency ($\nu_L$), the Einstein
coefficient $A_{u\ell}$ and the
collisional rates ($C_{ii^{\prime}}$), respectively. In addition, such
contributions change with time due to the MHD quantities ($\vec{\rm B}$, $\vec{\upsilon}_{\rm macro}$, T,
n) and to the radiation field tensor J$^K_Q$. As the atomic
rates are far larger than the temporal variations due to
macroscopic quantities, the partial temporal derivative in
Eq.(\ref{eq:ssev}) is relatively large and controlled by
atomic processes. Furthermore, the second term of the l.h.s.
can be disregarded against the time derivative, namely when
the amount of net $\rho^K_{Q,i}$ introduced in the plasma element at
  speed $\upsilon_{\rm{macro}}$ is
  negligible. In this situation, the
atomic collectivity reaches statistical equilibrium quickly. Hence,
Eq.(\ref{eq:ssev}) is usually solved by
\textit{forcing} the temporal derivative to zero and \textit{neglecting} the
spatial derivative (statistical equilibrium equation,
  SEE). This way, the
  atmosphere is considered stationary at intermediate time scales, the time
  step being assumed large enough to reach equilibrium
  ($\rm \Delta t >> t_{eq}$)
  but short enough to avoid significant net flow of material in the
  volume considered ($\rm \Delta t << t_{mat}$). The first condition is
  satisfied in our calculations\footnote{
 Non-equillibrium electron
densities and derived quantities are not discussed here because were
implicitly considered in the MHD modelling by accounting for partial
hydrogen recombination \citep{Carlsson:2016aa}.} for Ca {\sc i} $\lambda4227$ but it might fail in some atomic systems
  with long-lived metastable levels. The second
  condition has been assumed as valid in the present work but in
   chromospheric shock fronts the possible values of velocity,
  atomic polarization, density and collisions pose doubts
  requiring further investigation. 

Macroscopic dynamics is clearly key in the RTE.
 In a planar time-dependent stellar atmosphere with
 non-relativistic macroscopic speeds, the general RTE for the Stokes vector
${\bf\rm{I}}^0 (\vec{\Omega} ,x^0,s,t)=\rm (I,Q,U,V)^{\rm{T}}$ in a
reference frame comoving with 0 is
\begin{equation}
\centering
\frac{1}{c}\frac{\partial \rm{\bf{I}}^0}{\partial t}+\frac{\partial \rm{\bf{I}}^0}{\partial s} -\frac{\rm{d}
  \mathcal{V}}{\rm{d}s}\frac{\partial \rm{\bf{I}}^0}{\partial x^0} =
{\mathbf \epsilon}^0-{\bf K}^0 \rm{\bf{I}}^0, \qquad \textrm{}
\end{equation}\label{eq:eq1}
where all the symbols have standard meaning\footnote{Namely, s is the
geometrical distance along a ray $\vec{\Omega}=(\gamma,\eta,\mu)$
propagating at speed c with reduced frequency
$x^0=(\nu^0-\nu_0)/\Delta\nu^0_{\rm D}$ as seen by the plasma element
0, this latter having a thermal Doppler width $\Delta\nu^0_{\rm D}$
around the atomic transition frequency $\nu_0$, an emissivity 
$\mathbf{\epsilon^0}_{4\times1}$ and a propagation matrix
$\mathbf{K^0}_{4\times 4}$.} and $\mathcal{V}(s)=\vec{\xi}\cdot\vec{\Omega}$
is the projection along ray $\vec{\Omega}$ of the macroscopic Doppler velocity $\vec{\xi}$ (\textit{always} in
Doppler units at 0). Namely,
$\xi=\upsilon_{\rm{macro}}/(\upsilon^0_{\rm{thermal}}+
\upsilon^0_{\rm{turb}})^{1/2}$ is a ratio between resolved and
unresolved velocities.
Eq. (\ref{eq:eq1}) shows succinctly that the emergent solar polarization
depends on three very important points: (i) the spectral
structure of the radiation field seen from the scatterers
($\partial\rm{I^0}/\partial x^0$), (ii) the gradients of macroscopic
motions ($\rm{d\mathcal{V}/\rm{ds}}$) modulating such an incident
field, and (iii) its temporal
variation ($\partial\rm{I^0}/\partial
t$). In the comoving frame (CMF) RTE, macroscopic motions only appear in
an explicit dedicated term (third one in the l.h.s.), with no role anywhere else.
 Thus, such a CMF term is more sensitive to $\upsilon_{\rm{macro}}$ when
 the plasma is cooler ($\upsilon^0_{\rm{thermal}}\downarrow$), but
 vanishes when $\upsilon_{\rm{macro}}$ is constant or unresolved. It also shows that assuming
$\mathcal{V} \propto \mu\cdot \upsilon_z/\upsilon^0_{\rm{thermal}}$, as implicit when realistic models are treated in $1.5$D, just implies that vertical velocity \textit{gradients} are assumed much stronger than horizontal ones.
This describes a quiet chromosphere where weak magnetic
fields cannot guide shock waves or
gravity-accelerated flows towards the horizontal. In compliance
with the dominance of vertical variations, $1.5$D
calculations as ours consider an azimuthally-independent radiation
field, instead of the more general one affected by horizontal inhomogeneities
\citep[e.g.,][]{Stepan:2016aa}.
 
Note also that the l.h.s. of Eq. (\ref{eq:eq1})
is valid in uniform motion \citep{Mihalas:1978}: the acceleration of 0 between timesteps must be 
significantly smaller than the relative velocity between adjacent
plasma elements. This seems fulfilled in chromospheric models, also in shock waves. Finally, the
temporal variation can be treated implicitly by solving the whole problem
independently for each snapshot of the MHD simulation, which implies
the very good approximation $c^{-1}\partial{\bf I}/\partial t=0$.

 The solution to Eq. (\ref{eq:eq1}) for each ray and point in the
 atmospheric volume is ${\bf\rm{I}}^0 (\vec{\Omega},x^0,s,t)$. Such 
radiation field is shaped angularly by Doppler shifts \textit{along rays 
connecting} each location with the scatterer $0$. Thus, chromospheric
motions can efficiently generate anisotropic radiation that modulates
the polarization properties of the plasma and the emergent Stokes
vector in short temporal scales \citep{Carlin:2013aa}. 
%\articlefiguretwo{carlin_1_fig2a.png}{carlin_1_fig2b.png}{fig:fig2}{Temporal variation.}
\begin{figure}[t!]
\centering$
\begin{array}{cc} %[scale=0.4] [width=2.0in] --> options to control size
\includegraphics[scale=0.45]{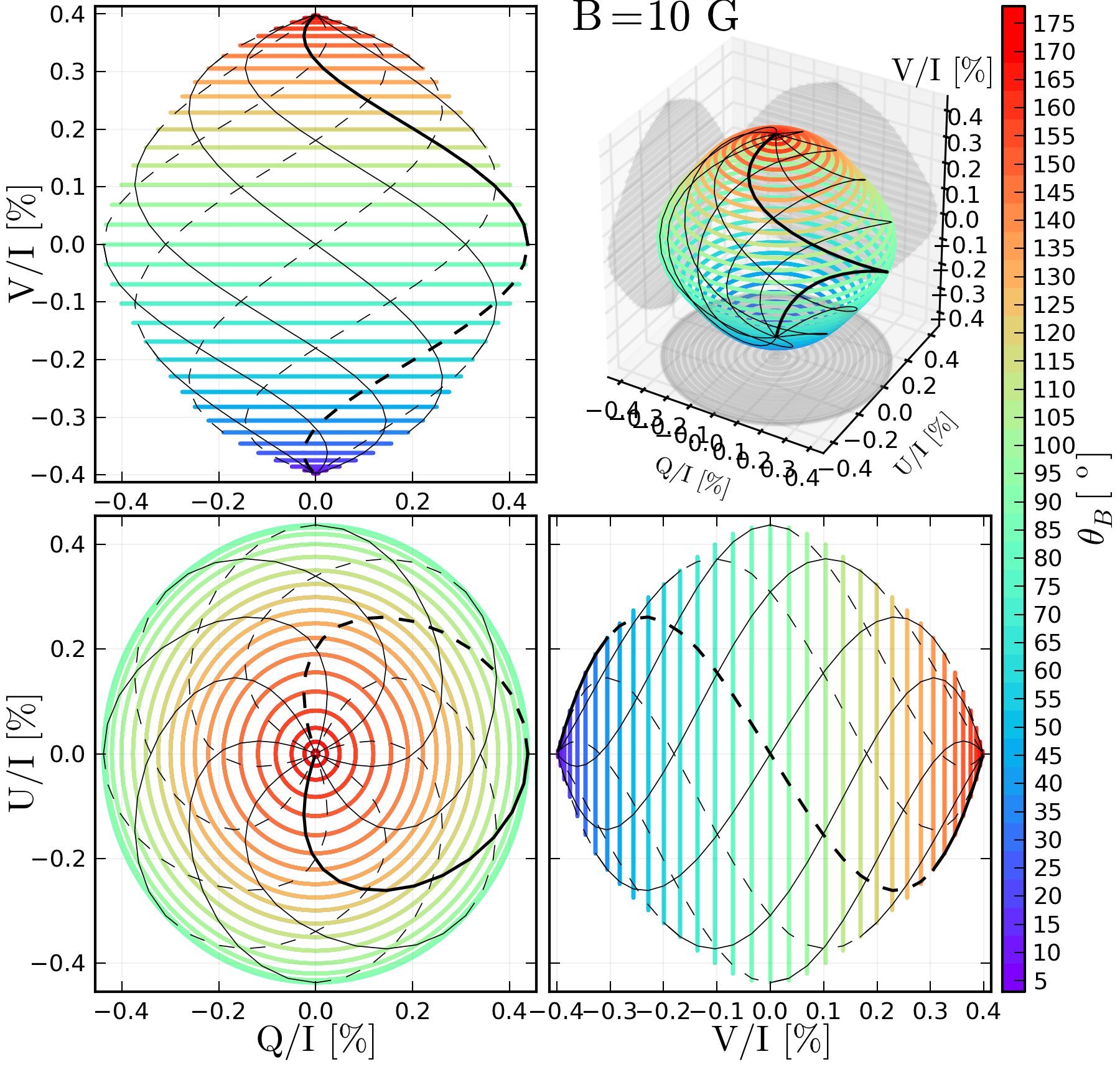} &\hspace*{-1.0em}
\includegraphics[scale=0.45]{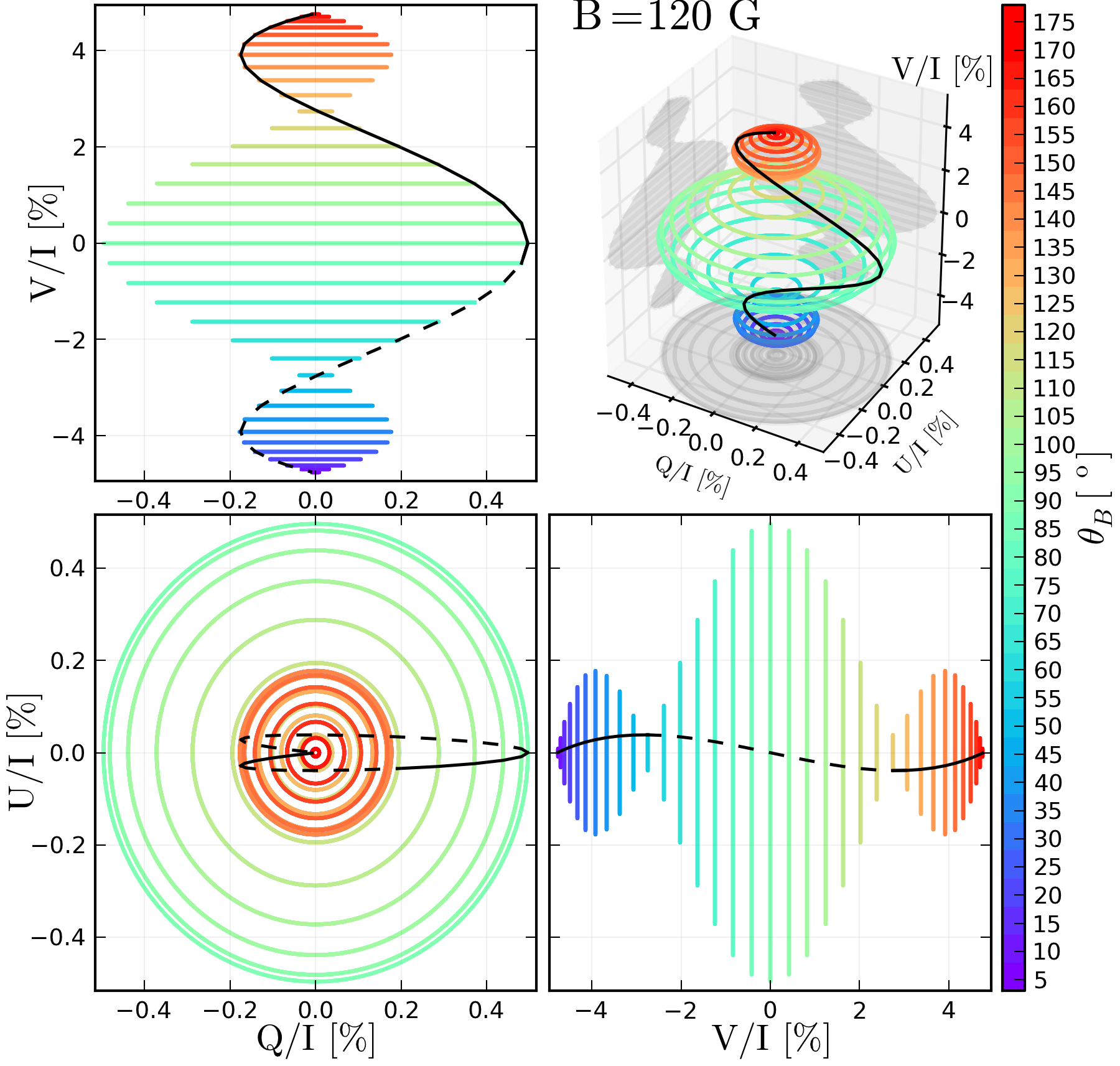}  
\end{array}$
\caption{\footnotesize Poincaré diagrams in (static) semiempirical FALC models for
  ad-hoc B$=10$ G (left) and B$=120$ G (right) in $\mu=1$. In black: some lines
  of $\rm \chi_B=cte$.}
\label{fig:fig2}
\end{figure}
\section{Poincaré diagrams and Hanle effect in static models}\label{sec:poincare}
Before considering dynamic signals it
is useful to characterize the Hanle and Zeeman effects for the given spectral line. To do it, we propose to
represent the polarization in the space Q,U,V (Poincaré diagram), as in the Figure \ref{fig:fig2}.

In Ca {\sc i} $4227$
{\AA}, the upper-level Hanle critical field is $\approx 20$-$25$ G, which implies Hanle sensitivity to magnetic
fields between $5$ and $125$ G. 
Figure \ref{fig:fig2} shows two main effects of increasing 
the magnetic strength in that range at disk center. The first one is the progressive
cancellation of the linear polarization (LP) as the magnetic field inclination $\theta_{\rm
  B}\rightarrow 54^{\rm o},125^{\rm o}$ (Van-Vleck angles). This
effect is maximum in full Hanle saturation (right panel) and can
produce Hanle polarity inversion lines \citep[HPIL,][]{Carlin:2015aa}
in synthetic maps of scattering polarization. Calculating for other field
strengths we find  
that Van-Vleck HPILs are already quiet developed (LP<LP$^{\rm max}$/4) for
B$\geq60$ G, hence full Hanle saturation is not necessary to create
such features in a spatial map. Poincaré diagrams reveal
a second interesting aspect of the forward-scattering Hanle effect. A same magnetic field azimuth produces
different U/Q ratios depending on the magnetic field
inclination. Then, the spectropolarimetric
 azimuth can be estimated (also out of
 saturation) with $\rm\tan\chi_B=\,$U/Q if subtracting an
 inclination-dependent magnetic field azimuth
\textit{offset} given by V/I (see forthcoming paper).
\begin{figure}[t!]
\centering$
\begin{array}{c}
\includegraphics[scale=0.93]{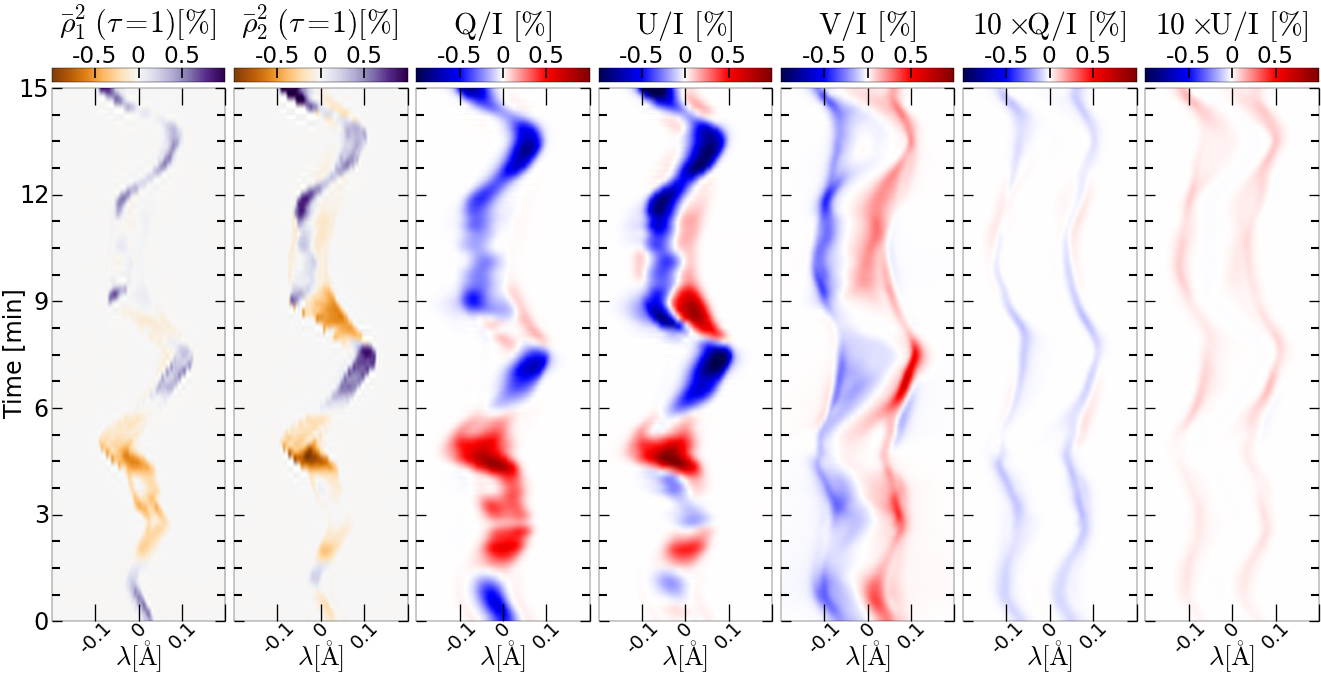}\\
\includegraphics[scale=0.75]{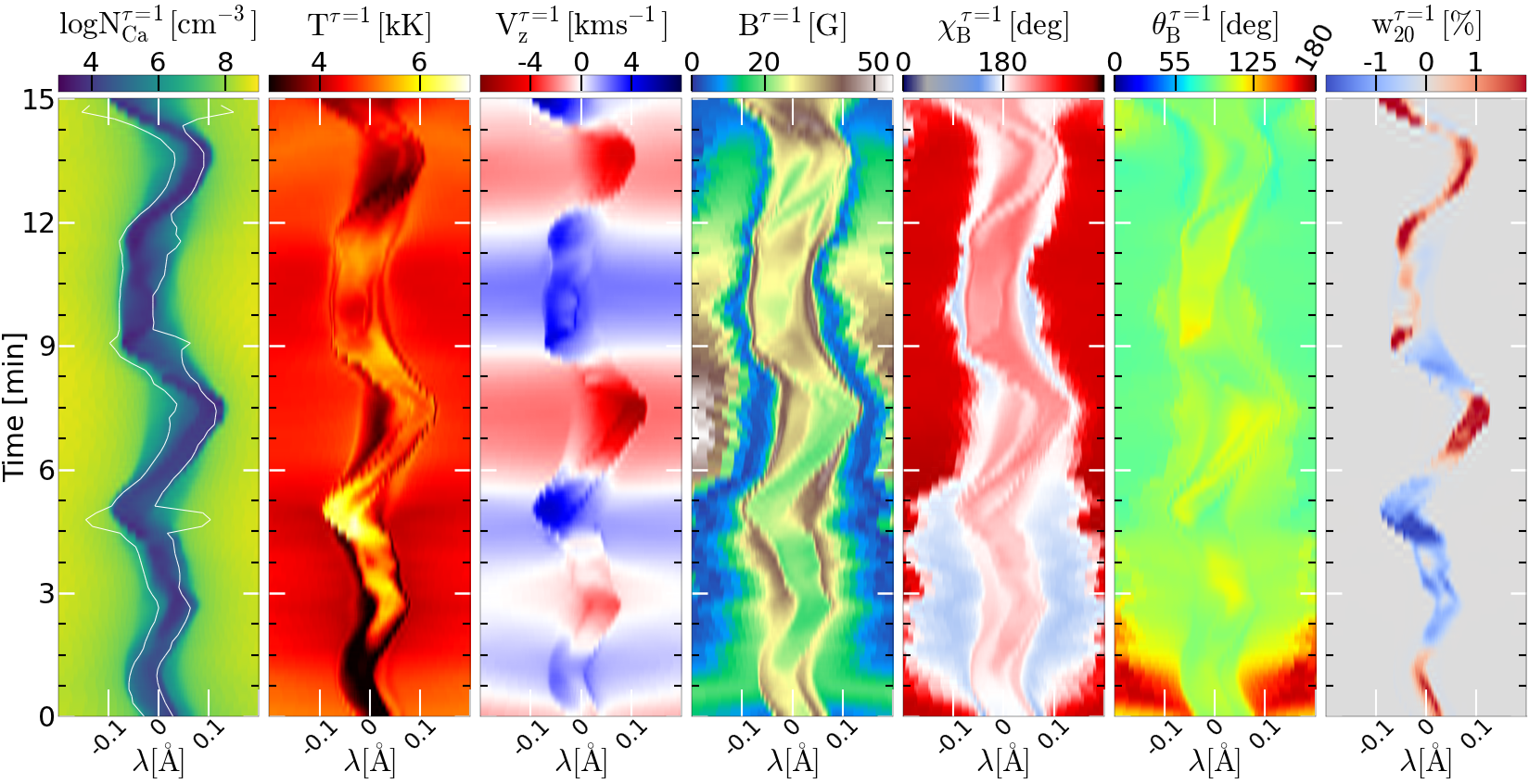}  
\end{array}$
\caption{\footnotesize Time serie in one pixel of the slit. Upper
  panel: real parts of $\rho^2_1$ and
  $\rho^2_2$ (lambdafied at 1), Q/I and U/I Hanle, V/I, $10\times$Q/I and $10\times$U/I Zeeman. Lower panel: lambdification at $1$ of atmospheric
  quantities. $\rm w_{20}$ is radiation field anisotropy.}
\label{fig:fig3}
\end{figure}
\begin{figure}[h!]
\centering$
\begin{array}{cc}
\includegraphics[scale=0.55]{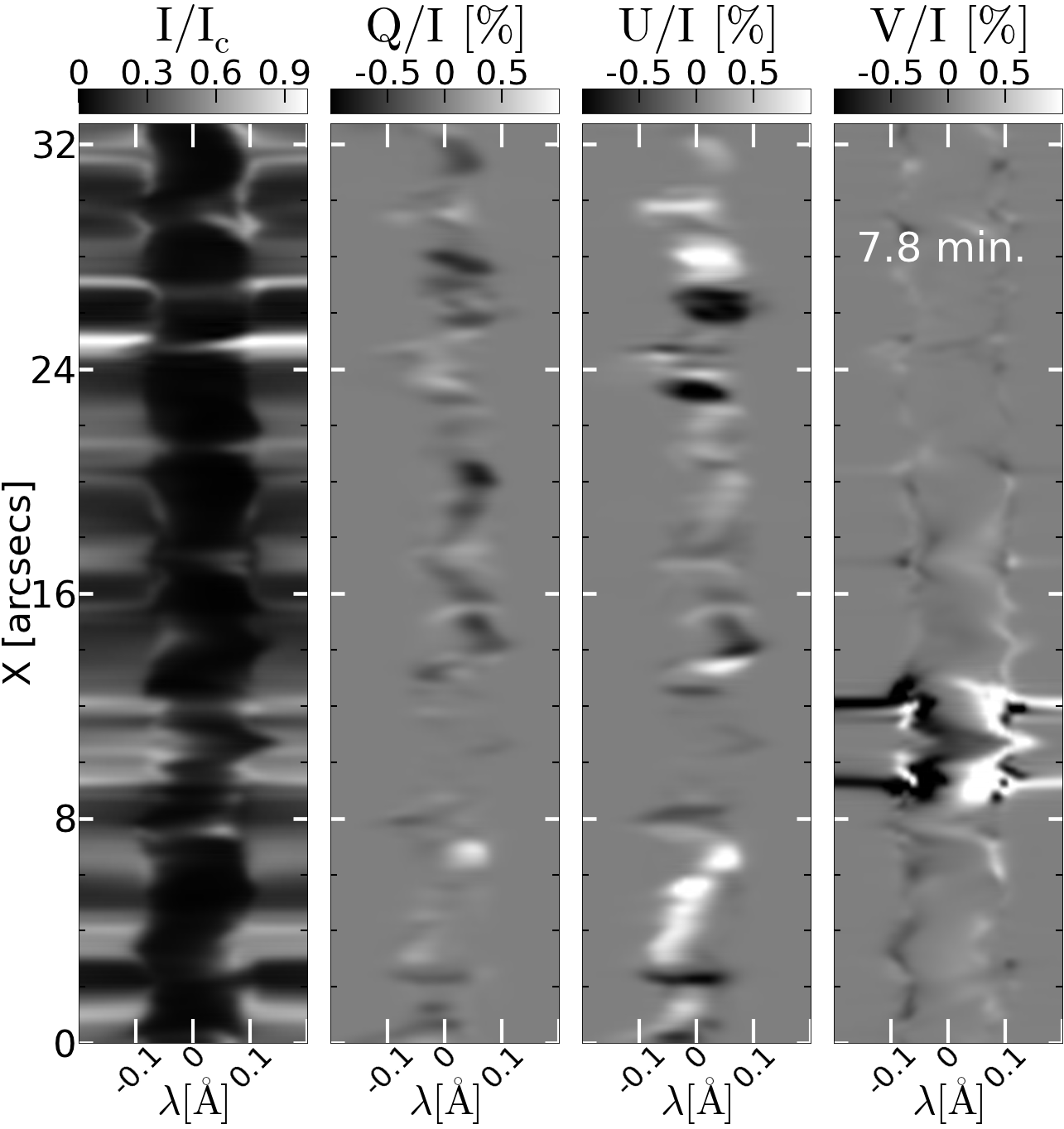} &
\includegraphics[scale=0.5]{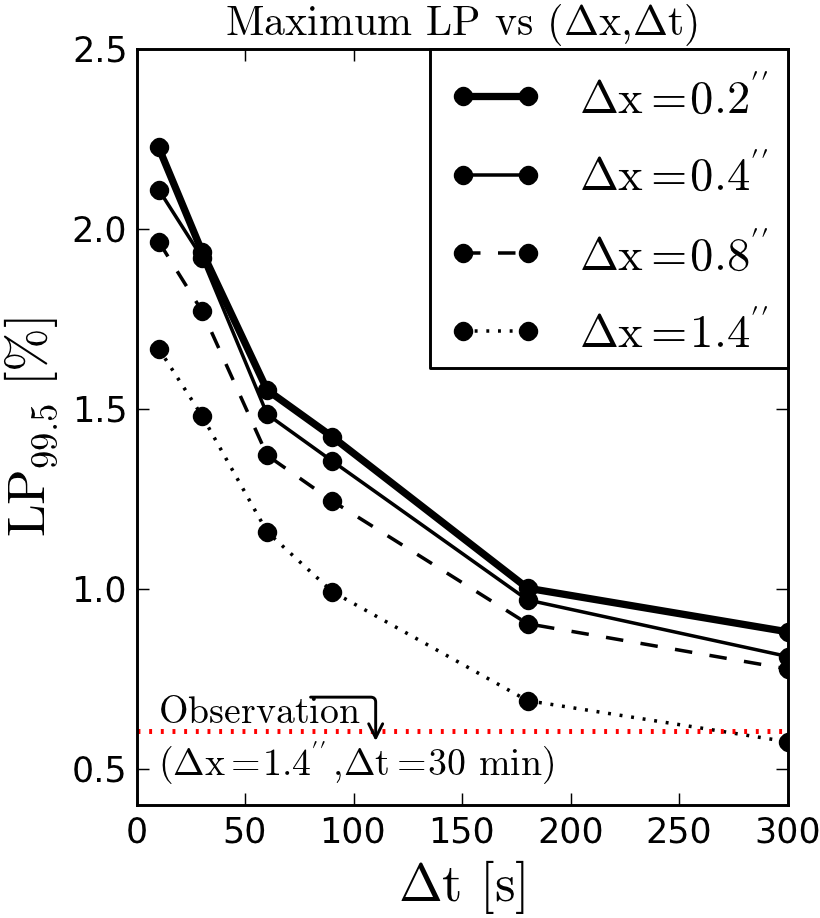}  
\end{array}$
\caption{\footnotesize Left: synthetic slit profiles in $\rm t=7.8$ min
  (after integrating the width of the slit). Right: dependence of the maximum total LP (percentile 99.5\%) on resolution ($\rm \Delta x$,$\rm \Delta t$). }
\label{fig:fig4}
\end{figure}
%\articlefigure{carlin_1_fig1d.png}{fig:fig1d}{Temporal variation.}
\section{The dynamic Hanle effect of Ca {\sc I} 4227{\AA}}
To analyze the temporal evolution, the transformation \textit{lambdafy
  M at} $\alpha$ is defined as $\rm M_{xy}(z,t) \rightarrow
 M_{x,y}(\lambda_{\tau=\alpha},t)$. Thus is how Fig. \ref{fig:fig3} 
associates quantities at $\tau(\lambda)=1$ with Stokes
profiles\footnote{Note that for showing information in heights above
  $\tau=1$ we would need to lambdafying at $\tau<1$. The complementary operation would be to get M for
  all $\tau$ at fixed $\lambda$, instead of obtaining M for all
  $\lambda$ at fixed $\tau$.} in $\lambda$. The figure shows the temporal
evolution of synthetic Hanle and Zeeman polarization
in one pixel at quiet sun for
$\lambda 4227$. Note the much weaker transversal Zeeman, the
interesting spectral variability (symmetries,
shifts) and the amplitudes of the Hanle LP along time. A reason for the
large LP amplitudes is that the magnetic strength and inclination vary close to values ($\theta_B=90,\,B\approx22 $G)
maximizing the forward-scattering Hanle effect in the chromospheric line
core (see Figs.\ref{fig:fig2} and \ref{fig:fig3}). Additionally, the
LP is substantially larger than in
observations because the
spatio-temporal resolution is increased, hence also the effect of
the parameter
$\xi$ introduced in Eq.(\ref{eq:eq1}) and controlling the LP amplitudes through
velocity gradients and radiation field anisotropy, as explained in
\cite{carlin12}.  Indeed, time integration can reduce the LP substantially, as seen in Fig.\ref{fig:fig4}, where the response of the maximum LP values in
our synthetic slit is shown for different resolutions. Furthermore, it was found that the spectral variability combined
with integrations larger than $3$ minutes changes
completely the morphology of the signals.
As these modulation effects are line
dependent, \cite{Carlin:2016aa} have pointed out that temporal
evolution/integration with macroscopic motions can account for the anomalous
excesses of line-core LP in the second solar spectrum that have
puzzled our community for years \citep{Stenflo:2000}. 

Unfortunately, measuring time evolution is hard. A decent
S/N usually implies time integration because solar
 scattering polarization is weak (hence noisy). 
But our results show that part of that weakness might be due to a
combination of dynamic issues \citep{Carlin:2016aa}. One of them is the
lack of temporal resolution, which partially cancels the LP because the
latter can change of sign in a single profile
as well as between close timesteps and pixels. Thus, at non-full
 resolutions, cancellation makes LP more sensitive to noise and
even longer integrations seem necessary. But, what happens when a
sensitive spectropolarimeter is combined with a full
resolution avoiding cancellations?
 
Observations with the ZIMPOL camera \citep{Ramelli:2010aa}
having poor spatial resolution
($0.^{\prime\prime}6-1.^{\prime\prime}4$), good temporal resolution
($\rm t_{integ}<30$ s) and medium effective integration ratio ($\rm r_{eff}= t_{eff}/t_{integ}\approx
 0.55$) \textit{give}
 \textit{large noisy} signals. Therefore, the amplitude that might
 have been gained by resolving in time the LP (and its enhancements produced by
 motions) is not enough against the intrinsic noise at such poor spatial
 resolution and medium effective integration time.  
What are the keys for going
 further? First, the instrumental time
 scales (see Fig. \ref{fig:timeline}). While $\rm t_{integ}$ has to be
minimized, the net temporal integration $\rm t_{eff}$ has to
 be maximized (larger $\rm r_{eff} $), which is limited by
 the readout time of the camera and by the 
 \textit{exposure} time ($\rm t_{exp} $) that saturates the pixel in
 intensity.
The second key is noise management. The design maximizing $\rm r_{eff}
$ (e.g., higher sampling/shorter $\rm t_{exp}$ or viceversa) needs also to consider and reduce the kind
of noise dominating during exposure time. The existence of detection suggests that photon noise
 dominates in our case (thermal noise is not a problem in the
 violet for cooled CCDs). Finally, full spatial resolution is required, meaning a large telescope with a stabilization system working in on-disk quiet-sun, which is also very challenging.
 
In summary, measuring polarization in chromospheric scales requires well-known solutions:
 (i) minimize noise,  by increasing the detected intensity with larger telescope apertures
 and efficiencies, by maximizing the \textit{net} temporal
 integration, and by a suited lower-noise camera design. And (ii), increase
 spatiotemporal resolution, again with large telescopes and by optimizing the temporal
 processes in the camera. The achievement of top
spatio-temporal resolution ($<15$ s, $<0.2^{\prime\prime}$) should expose intrinsically
larger LP Hanle signals by avoiding cancellations (at least in
$\lambda4227$), hence
improving our ability to discern the real (time-resolved) second solar spectrum. 

\section{Conclusions}
Temporal evolution is key for a true understanding of
the second solar spectrum and for 
diagnosing chromospheric magnetic fields. First, because morphological (spectral)
information is lost without it, but also because temporal resolution
itself avoids spectral cancellations that Hanle polarization
might be prone to suffer in moving media. Thus, LP amplitudes can increase, and
magnetic fingerprints encoded in null-polarization lines (Hanle PILs,
see Sec.\ref{sec:poincare}) might be more easily
detected by spatial contrast. 

Consequently, the modelling of polarization should consider dynamics (time evolution and macroscopic
motions) when possible. Further investigation is needed for
assessing how important are 
the streaming terms in the SEE and the subsequent possible lack of statistical
equilibrium in shock waves.

 On the other hand, Hanle time series at
chromospheric time scales demand large telescopes and/or an exquisite
management of noise and temporal efficiency in the spectropolarimeter.
These challenges require more research and
technological advances, but they are key (and achievable) steps for providing with 
effective Hanle diagnosis techniques to the solar community.
\\
\\
\acknowledgements The author thanks Michele Bianda, Daniel Gisler and Renzo Ramelli
for their support and feedback in regard to observational and
instrumental matters. This work was financed by the SERI project C12.0084 (COST action
MP1104) and by the Swiss National Science Foundation project $200021$\texttt{\_}$163405$.

%\bibliography{../../../mybibdesk_1}

% % For non-BibTex:
% \begin{thebibliography}{}
% \bibitem[Barnes (2008)]{ex_1}
% The first reference.  This reference may span the width of the page and should be in the format described in the instructions.
% \bibitem[Barnes (2009)]{ex_2}
% The second reference.  This reference may also span the width of the page and should be in the format described in the instructions.
% \bibitem[Barnes (2010)]{ex_3}
% The third reference.  If there is a URL in here make sure to put it in the right way.\\
% See {\footnotesize \url{http://www.somewhere.com/see_there's%still_characters_here}}
% \end{thebibliography}

\end{document}